# Strong linewidth variation for spin-torque nano-oscillators as a function of in-plane magnetic field angle

K. V. Thadani[1], G. Finocchio[2], Z.-P. Li[1], O. Ozatay[1], J. C. Sankey[1], I. N. Krivorotov[3], Y.-T. Cui[1], R. A. Buhrman[1], and D. C. Ralph[1]

[1] *Cornell University, Ithaca, NY 14853, USA*

[2] *University of Messina, Messina, Italy*

[3] *Department of Physics and Astronomy, University of California, Irvine, California 92697, USA*

## Abstract

We measure the microwave signals produced by spin-torque-driven magnetization dynamics in patterned magnetic multilayer devices at room temperature, as a function of the angle of a magnetic field applied in the sample plane. We find strong variations in the frequency linewidth of the signals, with a decrease by more than a factor of 20 as the field is rotated from the magnetic easy axis to the in-plane hard axis. Based on micromagnetic simulations, we identify these variations as due to a transition from spatially incoherent to coherent precession.

PACS:   75.75.+a, 72.25.-b, 76.50.+g

In a magnetic multilayer device, spin-transfer torque from a spin-polarized direct current can excite steady-state magnetic precession [1,2], thereby creating a nanoscale frequency-tunable microwave source [3- 17]. For applications, it is desirable that the microwave signal has a frequency spectrum with a narrow linewidth at room temperature. For this reason, understanding what physical processes affect the linewidth has generated considerable interest, both theoretically [18- 22] and in experiments exploring the dependence on temperature [12,13] and on magnetic fields which rotate the precession axis out of the sample plane [5-7,9,15]. Here we report measurements and simulations of a surprisingly strong dependence of the linewidth on the in-plane angle of applied magnetic field. We find that the most-commonly studied field orientation, in-plane and parallel to the magnetic easy axis, produces the broadest linewidths. As the field angle is rotated towards the in-plane hard axis, the linewidths decrease dramatically, by more than a factor of 20 in some devices. Comparisons with micromagnetic simulations suggest that this change is due to a crossover from spatially incoherent to coherent dynamics.

We will report results from two types of multilayer devices, both with a nanopillar geometry [3]. In both cases the magnetic "free layer" that precesses is 4 nm of permalloy (Py, $Ni_{81}Fe_{19}$). In the first geometry, the magnetic "fixed layer" that polarizes the current is 4 nm of Py exchange-biased to a layer of antiferromagnetic IrMn. The full layer structure is (with thicknesses in nm) 4 Py / 80 Cu / 8 IrMn / 4 Py / 8 Cu / 4 Py / 20 Cu / 30 Pt and a Cu top contact. The second type of sample has a thicker Py fixed layer (20 nm) with no exchange bias: 2 Py / 120 Cu / 20 Py / 12 Cu / 4 Py / 12 Cu / 30 Pt with a Cu top contact. The layers are deposited by sputtering, and electron-beam lithography and ion milling are used to etch through to the bottom Cu contact, giving a device cross section that is approximately elliptical (Fig. 1a inset) [23]. We will show data from one 50 × 150 nm² exchange-biased fixed-layer device (with the exchange bias directed along the long axis of the ellipse) and one 70 × 130 nm² thick-fixed-layer device. Similar results were obtained in 17 exchange-biased samples and 5 thick-fixed-layer devices.

Figure 1 shows the differential resistance (*dV/dI*) as a function of current (*I*) and magnetic field (*H*) for an exchange-biased device at room temperature. In agreement with previous measurements [3], as the current is swept (Fig. 1(a)), at low magnetic fields we observe hysteretic switching between the parallel and anti-parallel magnetic orientations, and above $H = 450$ Oe we find non-hysteretic peaks in *dV/dI* that are associated with transitions among precessional and static magnetic states. From the magnetic field dependence (Fig. 1(b,c)) we can infer that the effective dipole field of the fixed layer acting on the free layer is $H_{dipole} = 80$ Oe, the coercive field of the free layer at room temperature is $H_{an} \approx 200$ Oe and the sum of the exchange bias and effective field on the fixed layer is $H_{eb} = 360$ Oe. From a similar analysis for a thick-fixed-layer device, we find $H_{dipole} = 400$ Oe and $H_{an} \approx 100$ Oe for the free layer, with no exchange bias.

We measure [3] the high-frequency voltage oscillations produced by spin-torque-driven magnetic precession at room temperature, with a large in-plane magnetic field applied using a projected field electromagnet [24] that allows us to control the field angle $\theta_H$ continuously. For the field values used in our experiment (800-1000 Oe), both the measured resistances and macrospin modeling using the parameters determined above indicate that for the exchange-biased samples the offset angle between the magnetic moments of the two layers grows from 0º to ~35º as $\theta_H$ is increased from 0º to 90º, while for the thick-fixed-layer devices without exchange bias the offset angle is always < 5º.

Figures 2(a) and 2(b) show DC-driven spectra for the exchange-biased device at 1000 Oe and the thick-fixed-layer device at 800 Oe, at selected values of $\theta_H$. The current bias is 5 mA, significantly larger than the critical current needed to excite magnetization dynamics for any $\theta_H$. In the exchange-biased device, when $H$ is applied along the magnetic easy axis in the direction of the exchange bias ($\theta_H = 0°$), there is no visible precessional peak, just a low-frequency tail indicating aperiodic dynamics. A precessional peak is first resolvable for $\theta_H \approx 25°$, and at $\theta_H = 45°$ there is a broad peak near 6 GHz with a linewidth (full width at half maximum, FWHM) of 2 GHz. As $\theta_H$ is increased further, the linewidth decreases dramatically. The minimum linewidth for $I = 5$ mA is 170 MHz at $\theta_H = 95°$, a factor of 20 narrower than at $\theta_H \approx 25°$ (Fig. 2(c)). For the thick-fixed-layer device (Fig. 2(b,d)), a broad peak is visible even for $\theta_H = 0°$ at the second harmonic of the precession frequency. (Only a second harmonic signal is expected when the offset angle between the magnetic layers is zero [3].) For $\theta_H = 90°$, this second harmonic linewidth is reduced to 450 MHz, a factor of 5 less than at $\theta_H = 0°$.

In order to determine why the linewidths vary so strongly, we have analyzed the linewidth, precession frequency, and power of the precessional signals as a function of field angle and current for the two types of samples. For the exchange-biased-fixed-layer samples (Fig. 3), as a function of increasing field angle up to $\theta_H \approx 90°$ the signal displays a decreasing frequency and an increasing total power, together with the decreasing linewidth. The increasing power suggests that the precession amplitude grows as a function of $\theta_H$ at fixed current, while the frequency shift is consistent with an increasing demagnetization field and an increased precession amplitude. The narrowest linewidths are observed for $\theta_H$ between 75° and 95°, and for small currents (2.5 - 4 mA), the minimum linewidths approach 50 MHz, close to the resolution bandwidth employed in the measurements. Beyond $\theta_H \approx 95°$ (the exact value is current dependent), the total power in the precessional signal drops abruptly by a factor of 10 and the frequency undergoes changes in slope and jumps as a function of $\theta_H$. We suspect that these changes may be associated with transitions in the magnetization state of the fixed layer. The dependence of the power spectrum on $\theta_H$ (for $I = 4$ mA) is summarized in a logarithmic-scale plot in Fig. 3(d).

For the thick-fixed-layer sample (Fig. 4), the linewidths generally decrease with $\theta_H$ between $\theta_H = 0°$ and 90°, and increase between $\theta_H = 90°$ and 180° at all currents. In the ranges 50°-75° and 105°-150° the fits appear to suggest a non-monotonic dependence on $\theta_H$, with peaks and abrupt jumps, but these are likely just artifacts of the fitting procedure, associated with the fact that the spectra at these angles seem to consist of two closely-spaced peaks (see Fig. 4(d)) that are not well-described by Lorentzian fits. The measured frequencies vary smoothly as a function of $\theta_H$ (Fig. 4(b)), with a form that is approximately symmetric about $\theta_H = 90°$, as expected in the absence of any exchange bias. As noted already in Fig. 2, the decrease in the linewidth between $\theta_H = 0°$ and 90° for the thick-fixed-layer sample is less dramatic than the factor of 20 observed for the exchange-biased fixed layer samples.

We have performed both simple macrospin modeling and more detailed micromagnetic calculations in an attempt to understand these data. In the macrospin simulations, we explored a range of parameters chosen to approximate the sample characteristics and found some narrowing in the predicted linewidths as $\theta_H$ is increased from 0° to 90° at room temperature, but only by a factor of 2-3, not by the much larger factor observed experimentally. The macrospin simulations also predict that steady-state precession is stable over a much narrower range of current than we find experimentally.

Whereas macrospin simulations are unable to explain the large linewidth changes we measure, micromagnetic calculations give much better agreement. We performed simulations using the algorithms described in Ref. [25], which integrate the Landau-Lifshitz-Gilbert equation for both magnetic layers and include a Slonzcewski spin-torque term [26], the Oersted field from the current, the magnetic interaction between the layers, and fluctuating Langevin fields to model thermal fluctuations. Parameters are chosen to model the exchange-biased sample and are listed in the caption for Fig. 5.
Figure 5(a) shows examples of the simulated resistance as a function of time for an exchange-biased sample. The corresponding spectral densities of the resistance oscillations are shown in Fig. 5(b). We find that, due to thermal fluctuations at room temperature, there is no well-defined precessional peak in the simulated signal between $\theta_H = 0°$ and approximately 30°, consistent with our measurements. As $\theta_H$ is increased to 90°, the linewidth decreases strongly, eventually reaching a minimum (FWHM) of 350 MHz. The micromagnetic simulations suggest that the nature of the magnetization dynamics is qualitatively different for $\theta_H$ near 0° and 90°. In Figs. 5(d) and 5(e), we show snapshots of the spatial distribution of the magnetization in the free layer at $T = 300$ K for both field angles. The arrows in these plots represent the magnetization in the sample plane and the colors represent the magnetization component parallel to the long axis of the ellipse. For the case of $\theta_H = 0°$ (Fig. 5(d)), the oscillations are spatially non-uniform, and the left and right halves of the ellipse can process in opposite directions. In contrast, for the case of $\theta_H = 90°$ (Fig. 5(e)), the dynamics are nearly spatially uniform. For angles near $\theta_H = 45°$ the magnetization undergoes by ns-scale jumps between spatially nonuniform states and the state which is more spatially coherent.

Based on the micromagnetic simulations, we can consider the possible mechanisms underlying the crossover from spatially incoherent to coherent dynamics. One factor may be that the misalignment angle ($\theta_{mis}$) between the fixed and free layer moments grows as $\theta_H$ is increased in the exchange-biased samples. When $\theta_{mis}$ is much larger than the range of angular variations within the micromagnetic configuration of the free layer, the spin torque on each spatial element will be in the same direction, and this may promote spatially-coherent motion [25]. Another important factor may be the amplitude of the precession. In Fig. 5(a), we see that the amplitude of the oscillations is much smaller at $\theta_H = 0°$ and $\theta_H = 45°$ than at $\theta_H = 90°$. Smaller oscillation amplitudes make the dynamics more sensitive to thermal fluctuations in both amplitude and frequency [12]. A potential mechanism that does not appear to contribute to the crossover between spatially coherent and incoherent dynamics is the Oersted field from the applied current. We performed simulations with the Oersted field both included and absent, and found no significant qualitative differences in the dynamics.

In summary, we observe that applying a magnetic field along the hard in-plane axis of a magnetic nanopillar device, so as to offset the precession axis of the free layer away from the orientation of the exchange-biased fixed layer, can produce a dramatic reduction in the linewidth of the spin-torque-driven dynamics. Based on micromagnetic simulations, we associate this reduction with a crossover from spatially nonuniform magnetization dynamics to spatially coherent precession. This ability to control the spatial uniformity of the magnetization dynamics should help in the development of spin-torque nano-oscillators for use as microwave sources. Our results also suggest that large device-to-device variations in linewidths measured previously [3,8,9,12] may be associated with variations in the angle of the exchange bias relative to the field direction.

We thank B. Azzerboni, P. M. Braganca, C. Wang, P. G. Gowtham, and L. Torres for helpful discussions. We acknowledge support from the NSF/NSEC program through the Cornell Center for Nanoscale Systems, the Office of Naval Research, DARPA, and New York State. We also acknowledge NSF support through use of the Cornell Nanofabrication Facility/NNIN and the Cornell Center for Materials Research facilities.

**Figures & Figure Captions**

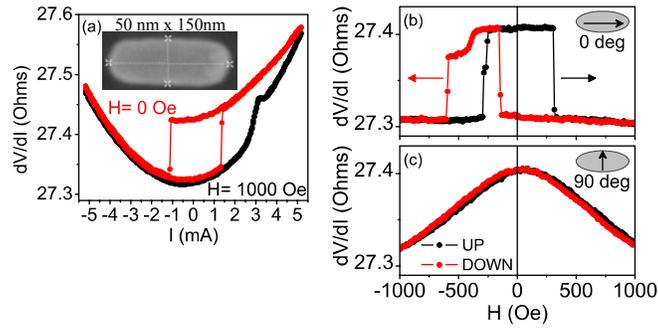

Fig. 1 (color online). (a) Differential resistance of a nanopillar spin valve device with an exchange-biased fixed layer as a function of current at room temperature. Inset: top-view electron micrograph of device shape. (b,c) Resistance as a function of magnetic field for the same device, for fields (b) along the easy-axis direction and (c) along the in-plane hard axis.

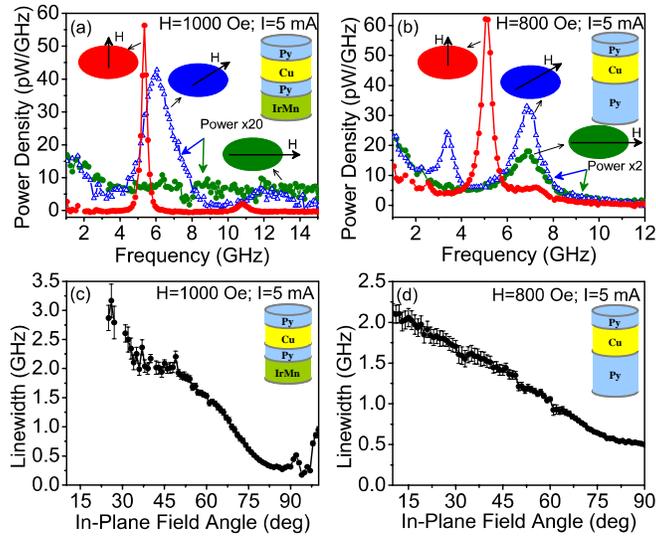

Fig. 2 (color online). Power spectral density of spin-torque-driven oscillations at room temperature, for field angles $\theta_H$ = 0º, 45º, and 90º (as labeled), for (a) an exchange-biased-fixed-layer sample and (b) a thick-fixed layer sample. Insets: sample structures. (c,d) Variation of the linewidth as a function of $\theta_H$ for both types of samples.

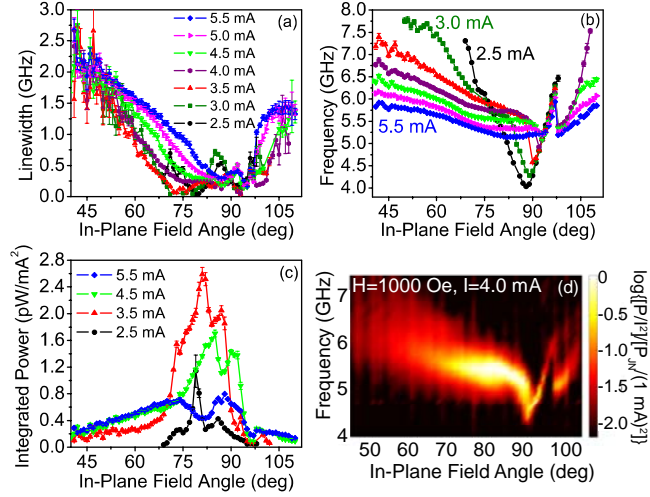

Fig. 3 (color online). Analysis of the spin-torque-driven microwave signals for the sample with the exchange-biased fixed layer, as a function of current and field angle at room temperature and $H$ = 1000 Oe. (a) Linewidth. (b) Peak frequency. (c) Integrated power within the precessional peak divided by $I^2$. (d) Power spectral density plotted on a logarithmic scale, as a function of $\theta_H$, for $I$ = 4 mA.

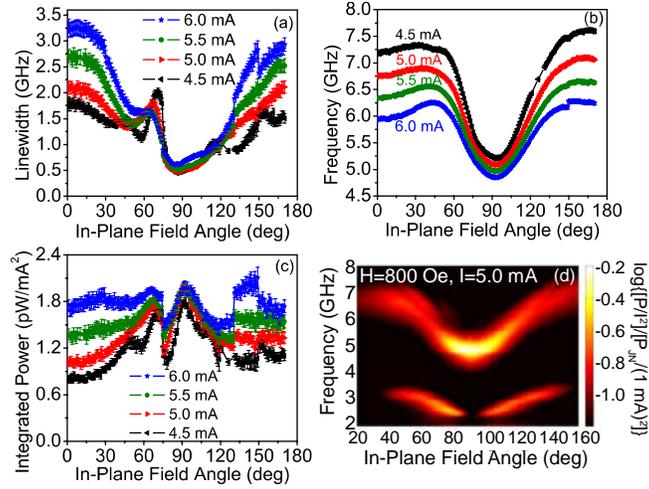

Fig. 4 (color online). Analysis of the spin-torque-driven microwave signals for the sample with the thick fixed layer, at room temperature and $H = 800$ Oe. (a) Linewidth. (b) Peak frequency. (c) Integrated power divided by $I^2$. (d) Power spectral density plotted on a logarithmic scale, as a function of $\theta_H$, for $I = 5$ mA.

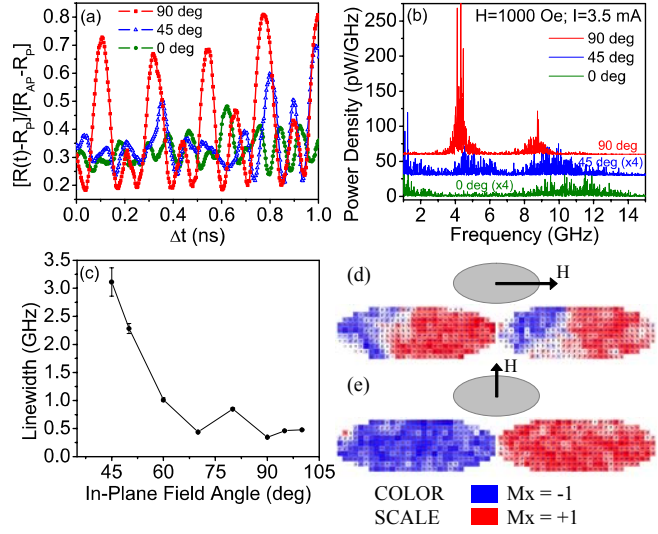

Fig. 5 (color online). Micromagnetic simulations of the exchange-biased sample. (a) Resistance as a function of time. $R_P$ is the resistance for parallel free and fixed magnetic layers, and $R_{AP}$ corresponds to antiparallel layers. (b) Corresponding power spectral densities. The curves for $\theta_H = 45°$ and $90°$ are offset by 30 pW/GHz and 60 pW/GHz. The curves for $\theta_H = 0°$ and $45°$ are scaled by a factor of 4. (c) Calculated linewidth as a function of $\theta_H$ for $H = 1000$ Oe, $I = 3.5$ mA. (d,e) Snapshots of the magnetization distribution in the magnetic free layer for (d) $\theta_H = 0°$ and (e) $\theta_H = 90°$, at times corresponding to maxima in the precession cycle. The parameters used in the micromagnetic simulations are: $H = 1000$ Oe, exchange-bias field, $H_{eb} = 360$ Oe, 20° relative to the easy axis; free and fixed layer saturation magnetization $M_S = 650$ emu/cm$^3$, free layer damping $\alpha = 0.025$; fixed layer damping = 0.2; temperature = 300 K, exchange constant = $1.3 \times 10^{-6}$ erg/cm, GMR asymmetry parameter = 1.5, current polarization = 0.38, and computational cell size = $5 \times 5 \times 4$ nm$^3$. Each simulation spans 100 ns with a time step of 0.334 ps.